\documentclass{article}
\usepackage{pdfpages}
\usepackage{xcolor}
\usepackage[mathlines]{lineno}
\usepackage{graphicx}
\usepackage{cite}
\usepackage{caption}
\usepackage{subcaption}
\usepackage{authblk}
\usepackage{amsmath}
\usepackage[top=5cm, bottom=5cm, left=3cm, right=2cm, headsep=1.5cm, heightrounded=true]{geometry}
\usepackage{fancyhdr}
\graphicspath{ {{images/}} }

\title{ Thermodynamic Topology of Black Holes within Tsallis	Statistics }
\author[1]{Yahya Ladghami\thanks{ \texttt{yahya.ladghami@ump.ac.ma}}}
\author[1,2]{Brahim Asfour\thanks{ \texttt{brahim.asfour@ump.ac.ma}}}
\author[1,2,3]{Amine Bouali\thanks{ \texttt{a1.bouali@ump.ac.ma}}}
\author[1,2]{Ahmed Errahmani\thanks{ \texttt{ahmederrahmani1@yahoo.fr}}}
\author[1,2]{Taoufik Ouali\thanks{ \texttt{t.ouali@ump.ac.ma}}}
\affil[1] {Laboratory of Physics of Matter and Radiation, Mohammed I University, BP 717, Oujda, Morocco}
\affil[2]{Astrophysical and Cosmological Center, BP 717, Oujda, Morocco}
\affil[3]{Higher School of Education and Training, Mohammed I University, BP 717, Oujda, Morocco}

\begin{document}
	\maketitle
	\begin{abstract}
	In this paper, we investigate the thermodynamic topology of black holes within the framework of Tsallis statistics. By integrating Tsallis non-extensive statistics with topological thermodynamics, we analyze the local and global stability of various black hole solutions, including Schwarzschild, Reissner--Nordström, and higher-dimensional black holes. The introduction of Tsallis entropy, parameterized by the non-extensive parameter, $\delta$, results in distinct thermodynamic behaviors depending on its value. Employing Duan’s $\phi$-mapping theory, we classify the thermodynamic topology of four-dimensional Schwarzschild black holes and non-charged higher-dimensional uncharged black holes into three distinct classes based on their topological number $W$: stable ($W = +1$), unstable ($W = -1$), and critical ($W = 0$). Additionally, the thermodynamic topology of Reissner--Nordström and charged higher-dimensional charged black holes is categorized into two classes, where $W = +1$ indicates a stable class and $W = 0$ represents a less stable class. Our study further demonstrates that the number of dimensions does not affect the topological thermodynamics within the context of non-extensive statistics. This approach provides novel insights into the interplay between Tsallis statistics and black hole thermodynamics, underscoring the pivotal role of topology in understanding black hole physics.
	\end{abstract}
\section{Introduction}
Black holes are among the most mysterious objects in the Universe. Moreover, understanding the
physics of black holes is crucial to addressing some of the most significant problems in theoretical
physics and science in general, such as the nature of quantum gravity, the unification theory, and the
information paradox. Black holes are a direct prediction of General Relativity, as their metrics are
solutions to Einstein’s equations. In General Relativity, black holes are defined as regions of spacetime
with extremely strong gravitational fields that prevent any form of matter or radiation from escaping.
Decades after their theoretical prediction, the existence of black holes was confirmed through the detection of gravitational waves from the merger of binary black holes~\cite{1} and the observation of the first
image of a black hole’s shadow~\cite{2,3,4,5}.
\\

The works of Stephen Hawking revolutionized our understanding of black holes. By considering
quantum effects, Hawking discovered that black holes emit blackbody radiation, now known as Hawking radiation~\cite{6}, which leads to the gradual evaporation of black holes. Furthermore, Jacob Bekenstein
demonstrated that black holes have entropy, which is proportional to their surface area rather than
their volume, as is the case for classical systems. This entropy is referred to as the Bekenstein-Hawking
entropy~\cite{7}. Building on the work of Hawking, Bekenstein, and Bardeen, the four laws of black hole
mechanics were formulated, analogous to the laws of traditional thermodynamics~\cite{8}. This framework is known as black hole thermodynamics.
\\

Significant studies on black hole thermodynamics have been conducted in Anti-de Sitter (AdS)
spacetime, which is characterized by a negative cosmological constant. In these studies, the cosmological constant is treated as a thermodynamic variable associated with the thermodynamic pressure~\cite{9}. This approach, referred to as “extended phase space thermodynamics”~\cite{10}, has greatly enhanced our understanding of black hole properties. Within this framework, phenomena such as phase
transitions~\cite{11,12,13,14}, critical behavior~\cite{15,16,17,18}, multi-critical points~\cite{19,20}, and Joule-Thomson expansion~\cite{21,22,23,24,25} have been explored and analyzed.
\\

Recently, a new thermodynamic approach has emerged based on the Anti-de Sitter/Conformal Field
Theory (AdS/CFT) correspondence~\cite{26}. This correspondence establishes an equivalence between the
thermodynamics of black holes in AdS spacetime and the thermodynamics of a dual Conformal Field
Theory (CFT) on the boundary. A precise dictionary connects black hole quantities, such as the
mass, entropy, and temperature, with corresponding quantities in the dual CFT~\cite{27}. This holographic
thermodynamics framework has deepened our understanding of the AdS/CFT correspondence and
the behavior of black holes. For instance, it has clarified the role of the central charge in black hole
stability~\cite{28,29,30,31}, and explored the critical phenomena exhibited by black holes~\cite{32,33,34}.
\\

To further understand black hole physics, Wei, Liu, and Mann~\cite{35} recently introduced the concept
of topology into black hole thermodynamics by treating them as topological defects. This approach
utilizes the generalized off-shell free energy and Duan’s topological current theory, also known as the
$\phi$-mapping method~\cite{36}. Through this framework, both the local and global thermodynamic stability
of black holes can be analyzed using specific topological quantities, namely the winding and the topological numbers. A positive winding number indicates stable black holes, while a negative winding
number corresponds to unstable black holes~\cite{35}. Using the topological number, black holes can be
classified into three main types, although some researchers argue that a fourth class may exist~\cite{37}. An
alternative approach has been proposed to study the topology of black hole thermodynamics, replacing
Duan’s topological current theory with the residue method. This method has been shown to produce
results consistent with those of Duan’s theory~\cite{38}. Thermodynamic topology is simple, robust, and
innovative, offering new tools for studying the properties of black holes and spacetime. It has been
applied in various works~\cite{39,40,41,42,43,44}, providing valuable insights into black hole thermodynamics.
\\

In the context of statistical frameworks, the Gibbs-Boltzmann statistics is not well-suited for studying black holes, as black hole entropy is non-additive. To address this problem, Tsallis and Cirto proposed a generalization of Gibbs-Boltzmann statistics~\cite{45}. This generalized framework is non-extensive
and includes a parameter that can be adapted to the system under study. Within this framework, black
hole entropy is modified as follows $S_T = (S_\text{BH})^\delta$, where $S_\text{BH}$ is the Bekenstein-Hawking entropy, and
$\delta$ is the non-extensive parameter.
\\

In the present work, we combine topological methods with Tsallis statistics to study the thermodynamic topology of black holes. This approach enables us to analyze both the local and global thermodynamic stability within a non-extensive statistical framework. Our goal is to explore how non-extensive statistics modify black hole thermodynamic properties and their corresponding topological characteristics. In this paper, we focus on various black hole solutions, including Schwarzschild, Reissner--Nordström, and both charged and non-charged higher-dimensional black holes, to study the impact of non-extensive statistics on their stability and classification. In particular, we examine the role of electric charge and the number of dimensions on the thermodynamic behavior within Tsallis statistics. Furthermore, it is well known that Schwarzschild black holes are always thermodynamically unstable under the Gibbs--Boltzmann framework, which makes them an ideal case to study the impact of different statistical frameworks on black hole thermodynamics. As we will show, Schwarzschild black holes in the context of Tsallis thermodynamics can exhibit both stable and unstable behavior. 
\\

The paper is organized as follows. In Section~\ref{SS1}, we introduce thermodynamic topology within the Tsallis statistical framework. In Section~\ref{SS2}, we apply this framework to study various black hole solutions.  In Section~\ref{SS3}, we extend our analysis to include higher-dimensional charged and uncharged black holes. In Section~\ref{SS4}, we present our discussions and conclusions. Throughout this paper, we adopt natural units where $\hbar = G = c = k_B = 1$.

\section{Thermodynamic topology via Tsallis statistics}
\label{SS1}
\subsection{Tsallis entropy}
In classical thermodynamics, the entropy of ordinary systems is proportional to the volume or size of
the system. However, for black holes, the entropy is proportional to the surface area as suggested by
Bekenstein rather than the volume. Furthermore, black hole entropy is non-additive, meaning that the
standard Gibbs-Boltzmann statistical framework is not suitable for studying black holes. To address
this issue, Tsallis proposed a generalization of the Gibbs-Boltzmann statistics, based on the concept
of non-extensive entropy.
\\

Tsallis entropy of black holes can be expressed as~\cite{45}

\begin{equation}
	S_T = \left(S_\text{BH}\right)^\delta, 
\end{equation}
where $S_\text{BH}$ represents the Bekenstein-Hawking entropy, and $\delta$ is a positive parameter characterizing
the degree of non-extensivity. In the case of $\delta = 1$, the standard Bekenstein-Hawking entropy is
recovered.
\\

Tsallis’s generalization is not limited to entropy alone but extends to other thermodynamic quantities. For instance, the temperature of a black hole, which is a conjugate quantity of entropy, can be
derived from the following relation
\begin{equation}
	T_T = \frac{\partial M}{\partial S_T}, 
\end{equation}
where $M$ is the black hole mass, and this temperature is known as the Tsallis temperature. Additionally,
the free energy within the Tsallis statistical framework can be expressed as
\begin{equation}
	\label{e3}
	F_T = M - T_T S_T. 
\end{equation}

The introduction of Tsallis statistics offers a broader framework for investigating black holes, allowing for a more comprehensive understanding of their thermodynamic properties. In the following
subsection, we will explore how the thermodynamic topology of black holes can be reconstructed within
the Tsallis statistical framework.
\subsection{Thermodynamic Topology}
We can study the thermodynamic properties of black holes using a topological approach, where black
holes are treated as topological defects. In this approach, we use the generalized off-shell free energy,
given by~\cite{35}
\begin{equation}
	\mathcal{F} = M - \frac{S}{\tau}, 
\end{equation}
where $M$ and $S$ represent the mass and entropy of black holes, respectively, and $\tau$ is the inverse temperature of the system. When $\tau$ corresponds to the black hole temperature, the free energy $\mathcal{F}$ becomes the
on-shell free energy.
\\

We find the generalized off-shell free energy within Tsallis statistics. Using Eq. \eqref{e3}, as follows
\begin{equation}
	\mathcal{F}_T = M - \frac{S_T}{\tau}, 
\end{equation}
where $S_T$ is the Tsallis entropy and for $\tau = \tau_T = \frac{1}{T_T}$ we recover the on-shell free energy within Tsallis
statistics. This generalization allows for the study of black hole thermodynamics in a broader statistical
framework, accommodating potential deviations from Boltzmann-Gibbs statistics.
\\

To analyze the thermodynamic topology of black holes, we employ Duan’s $\phi$-mapping theory. In this formalism, we
define a vector field $\phi$ as follows~\cite{35}
\begin{equation}
	\phi=\left(\phi^1, \phi^2\right)=\left(\frac{\partial \mathcal{F}_T}{\partial r_{\mathrm{h}}},-\cot \theta \csc \theta\right),
\end{equation}
where $r_h$ is the radius of the black hole’s event horizon, and $\theta$ is a parameter constrained within the
range $0 \leq \theta \leq \pi$. Furthermore, the topological current is defined by the following expression~\cite{35,36}

\begin{equation}
	\label{e7}
	j^\mu=\frac{1}{2 \pi} \epsilon^{\mu \nu \rho} \epsilon_{a b} \partial_\nu n^a \partial_\rho n^b,
\end{equation}
where $\mu, \nu, \rho=0,1,2, a, b=1,2$, and $\partial_\nu=\frac{\partial}{\partial x^\nu}$, with $x^\nu=\left(\tau_T, r_{\mathrm{h}}, \theta\right)$. Here, $n=\left(n^1, n^2\right)$ represents a unit vector defined as
\begin{equation}
	n^a=\frac{\phi^a}{\|\phi\|},
\end{equation}
where $\|\phi\|=\sqrt{\left(\phi^1\right)^2+\left(\phi^2\right)^2}$, is the norm of the vector field $\phi$. The topological current can also be
rewritten in terms of the vector Jacobi by the Dirac delta function associated with the vector field
$\phi(x)$, $\delta(\phi(x))$, as follows~\cite{36}
\begin{equation}
	\label{e9}
	j^\mu=\delta^2(\phi) J^\mu\left(\frac{\phi}{x}\right),
\end{equation}
It plays a crucial role in localizing the topological current at the zeros of $\phi(x)$. The vector Jacobi is
expressed as
\begin{equation}
	\epsilon^{a b} J^\mu\left(\frac{\phi}{x}\right)=\epsilon^{\mu \nu \rho} \partial_\nu \phi^a \partial_\rho \phi^b .
\end{equation}
From Eq. \eqref{e9}, that the topological current is equal to zero when $\phi^a\left(x^i\right)=0,$, where we denote the
i-th solution by $z_i$. From Eq. \eqref{e7}, one can easily show that the topological current is conserved. The
topological number\footnote{ In the literature $W$ is also denoted as the topological charge.} $W$ is constructed by means of this conserved current as follows~\cite{35}
\begin{equation}
	W=\int_{\Sigma} j^0 d^2 x=\sum_{i=1}^N \beta_i \eta_i=\sum_{i=1}^N w_i,
\end{equation}
where $\beta_i$ represents the positive Hopf index and counts the number of loops of the vector $\phi^a$, as $x^\mu$ goes around the zero point $z_i$ in the space of $\phi$. Additionally, $\eta_i$ is Brouwer degree, equals to $\operatorname{sign}\left(J^0(\phi / x)_{z_i}\right)= \pm 1$, and $w_i$ represents the winding number for the i-th zero point of $\phi$ in the domain $\Sigma$. The vector Jacobi can be expressed for $\mu=0$ using the following relation \cite{46}
\begin{equation}
	\label{e12}
	\begin{aligned}
		J^0\left(\frac{\phi}{x}\right) & =\frac{\partial \phi^1}{\partial x^1} \frac{\partial \phi^2}{\partial x^2}-\frac{\partial \phi^1}{\partial x^2} \frac{\partial \phi^2}{\partial x^1} \\
		& =\frac{\partial^2 \mathcal{F}}{\partial r_h^2}\left(\frac{1+\cos ^2 \theta}{\sin ^3 \theta}\right).
	\end{aligned}
\end{equation}
For the zero point $\phi = 0$, where the value of $\theta$ is equal to $\pi/2$, we simplify Eq. \eqref{e12} as follows
\begin{equation}
	J^0\left(\frac{\phi}{x}\right)=\frac{\partial^2 \mathcal{F}}{\partial r_h^2}.
\end{equation}
Now, we can find a simplified expression for the winding number at the zero point as follows
\begin{equation}
	\label{wn}
	w_i=\operatorname{sign}\left(\left[\frac{\partial^2 \mathcal{F}}{\partial r_h^2}\right]_{z_i}\right),
\end{equation}
where $z_i$ represents the solution of $\phi^1=0$.
\section{Topological Classification of Black Holes}
\label{SS2}
In this section, we investigate the thermodynamic topology of various black holes—including the Schwarzschild, Reissner–Nordström, and higher-dimensional black holes—by taking into account the non-extensive nature of black hole entropy via Tsallis statistics. 

\subsection{Schwarzschild Black Holes}
\begin{figure}[htp]
	\centering
	\includegraphics[width=0.55\linewidth]{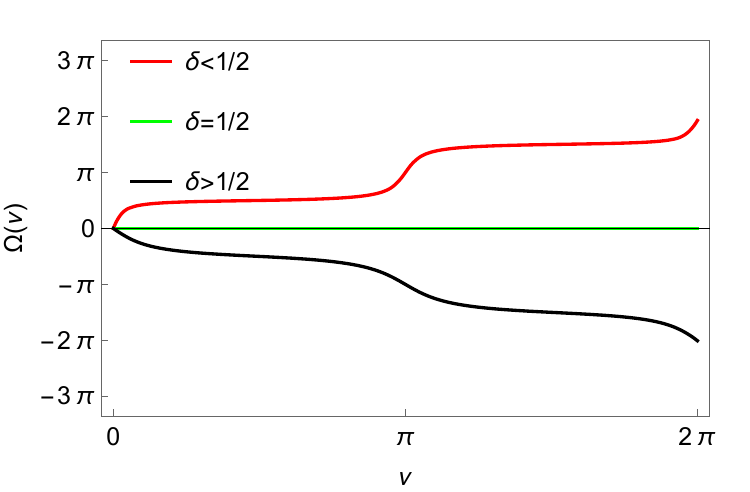}
	\caption{$\Omega-v$ curves for different values of the Tsallis parameter, where $a=b=0.4$ and $r_0=1$.}
	\label{f1}
\end{figure}

We explore the thermodynamic topology of Schwarzschild black holes within the framework of Tsallis statistics. The Schwarzschild black hole is described by the following metric
\begin{equation}
	d s^2=-\left(1-\frac{2 M}{r}\right) d t^2+\left(1-\frac{2 M}{r}\right)^{-1} d r^2+r^2 d \theta^2+r^2 \sin ^2 \theta d \phi^2,
\end{equation}
where $M$ is the mass of the black hole. The event horizon is located at $r_h = 2M$. In Tsallis statistics, the entropy of the Schwarzschild black hole is generalized to~\cite{45}
\begin{equation}
	S_T=\left(\pi r_h^2\right)^\delta.
\end{equation}
The Tsallis temperature $T_T$ of the Schwarzschild black hole is derived as
\begin{equation}
	T_T=\frac{\partial M}{\partial S_T}=\frac{r_h}{4 \delta\left(\pi r_h^2\right)^\delta}.
\end{equation}
To investigate the thermodynamic behavior and determine the topological invariants, we calculate the
off-shell free energy, given by
\begin{equation}
	\mathcal{F}_T=M-\frac{S_T}{\tau}=\frac{r_h}{2}-\frac{\left(\pi r_h^2\right)^\delta}{\tau},
\end{equation}
Furthermore, using the off-shell free energy, the components of the vector field can be expressed as
\begin{equation}
	\phi^1=\frac{\partial \mathcal{F}_T}{\partial r_h}=\frac{1}{2}-\frac{2 \pi^\delta\left(r_h^2\right)^\delta \delta}{r_h \tau},
\end{equation}
and
\begin{equation}
	\phi^2=-\cot \theta \csc \theta.
\end{equation}
The inverse temperature is parameterized as
\begin{equation}
	\tau=\frac{4 \delta\left(\pi r_0^2\right)^\delta}{r_0},
\end{equation}
where $r_0$ is an arbitrary positive parameter with the dimension of length. For $r_0=r_h$, the inverse temperature reduces to the Schwarzschild black hole's temperature in Tsallis statistics. Substituting the expression of $\tau, \phi^1$ becomes
\begin{equation}
	\phi^1=\frac{1}{2}-\frac{1}{2}\left(\frac{r_h}{r_0}\right)^{2 \delta-1}.
\end{equation}
The zero-point condition for the vector field $\phi$ is satisfied when $\theta=\pi / 2$ and $r_h=r_0$. To analyze the effect of non-extensive statistics on the thermodynamic topology of Schwarzschild black holes, we compute the winding number at this zero point using the following expression
\begin{equation}
	\label{e23}
	w=\operatorname{sign}\left[\left.\frac{\partial \phi^1}{\partial r_h}\right|_{r_h=r_0}\right]=\operatorname{sign}\left[\frac{1-2 \delta}{2 r_0}\right] .
\end{equation}
The winding number depends on the value of $\delta$. For $\delta<1 / 2, w=+1$ and for $\delta>1 / 2, w=-1$. Another method to determine the winding number involves using a contour around the zero point. The parametric equation of this contour is expressed as
\begin{equation}
	r=a \cos (v)+r_0, \quad \theta=b \sin (v)+\frac{\pi}{2},
\end{equation}
where the parameter $v \in[0,2 \pi]$. To find the winding number, we define a topological quantity that measures the deflection of $\phi$ along this contour given by the following expression \cite{47}
\begin{equation}
	\Omega(v)=\int_0^v \epsilon_{i j} n^i \partial_v n^j d v
\end{equation}
where $n^i$ are the components of the normalized vector field. This quantity is related to the winding number by the relation
\begin{equation}
	w=\frac{\Omega(2 \pi)}{2 \pi}
\end{equation}
Using this approach, we arrive at the same results as before: the topological number is related to the Tsallis parameter. Specifically, for $\delta<1 / 2, w=+1$; for $\delta>1 / 2, w=-1$; and for $\delta=1 / 2$, $w=0$. Additionally, Fig. \ref{f1} illustrates the evolution of $\Omega$ as a function of $v$ for all cases of the Tsallis parameter.
\\

\begin{figure}[htp]
	\centering
	\begin{subfigure}{0.3\textwidth}
		\centering
		\includegraphics[width=\linewidth]{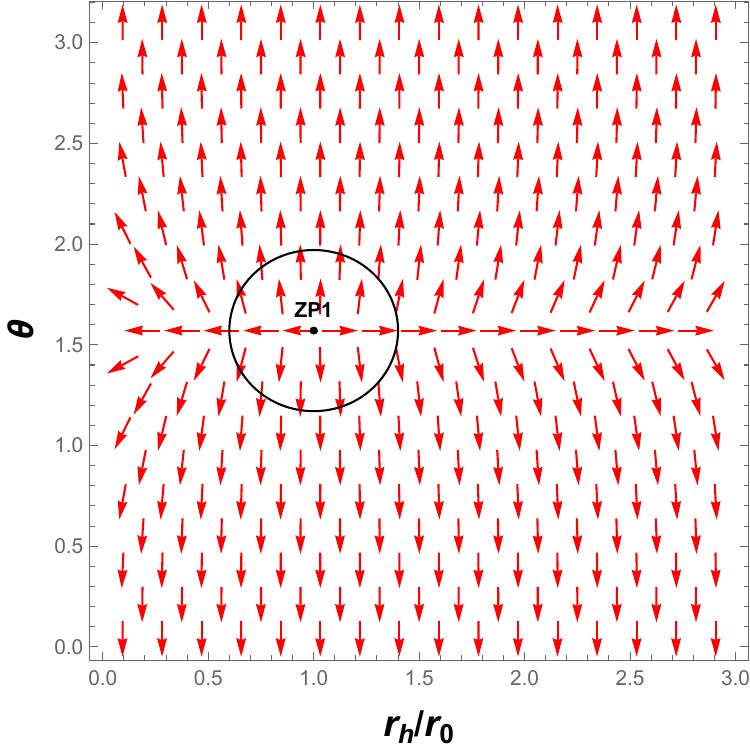}
		\caption{$\delta < 1/2$}
		\label{fig:a}
	\end{subfigure}
	\begin{subfigure}{0.3\textwidth}
		\centering
		\includegraphics[width=\linewidth]{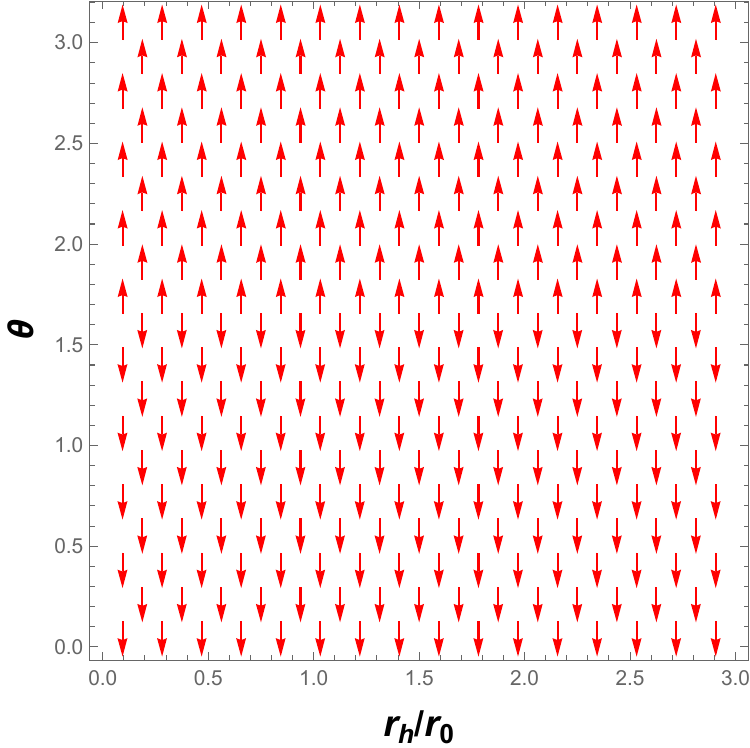}
		\caption{$\delta = 1/2$}
		\label{fig:b}
	\end{subfigure}
	\begin{subfigure}{0.3\textwidth}
		\centering
		\includegraphics[width=\linewidth]{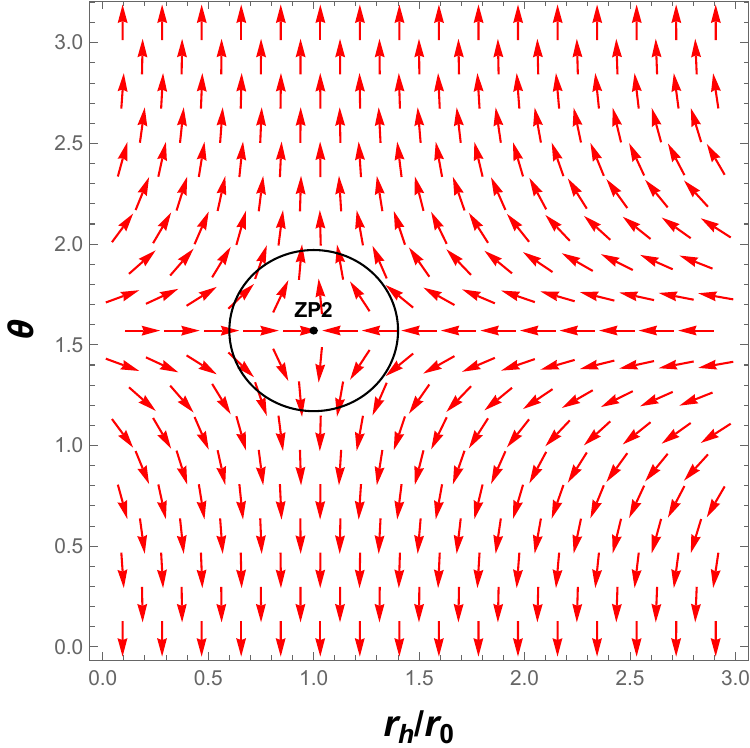}
		\caption{$\delta > 1/2$}
		\label{fig:c}
	\end{subfigure}
	
	\caption{ Vector field $\phi$ of Schwarzschild black holes for different values of the Tsallis parameter in $\theta-r_h / r_0$ plane.}
	\label{f2}
\end{figure}

Through Figs. \ref{f1}, \ref{f2} and Eq. \eqref{e23}, we investigate the impact of non-extensive statistics on the thermodynamic topology of Schwarzschild black holes. The analysis reveals three distinct scenarios,
each determined by the value of the Tsallis parameter $\delta$:

Case of  $\delta<1 / 2:$ The winding number associated with the zero points of the vector field is $w=+1$. This positive value reflects a topological defect indicative of local thermodynamic stability in the Schwarzschild black hole solution. For global topological properties, the topological number $W$, defined as the sum of the winding numbers of all zero points, is positive $W=+1$ in this case. This result demonstrates that the global thermodynamic behavior is stable when $\delta$ is less than the critical value $\delta_c=1 / 2$.

 Case of $\delta>1 / 2:$ The winding number becomes $w=-1$, indicating a topological defect associated with local thermodynamic instability. The topological number is negative, $W=-1$, i.e. a transition to a unstable thermodynamic phase when the Tsallis parameter exceeds the critical value.

Critical Case $\delta=1 / 2$ : At $\delta=1 / 2$, no zero points exist. This special value, referred to as the critical Tsallis parameter $\delta_c$, marks a boundary between distinct thermodynamic typologies. At this critical value, the system exhibits neither stability nor instability, and the topological number is found to be $W=0$.
\\

The aforementioned analysis classify the effect of Tsallis statistics on Schwarzschild black holes into three classes: $W=+1, W=0$, and $W=-1$, where each class exhibits a distinct thermodynamic behavior.
\\

\begin{figure}[htp]
	\centering
	\includegraphics[width=0.5\linewidth]{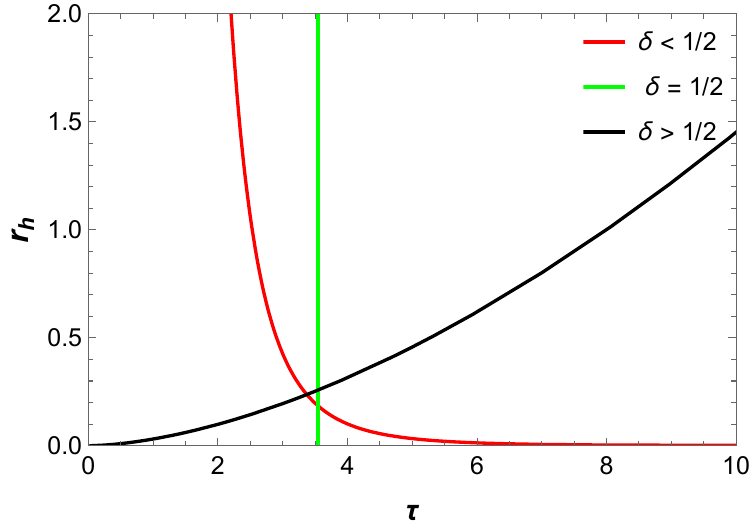}
	\caption{ Zero point curves of the vector $\phi^1$ in the $\tau-r_h$ plane for different values of the Tsallis parameter}
	\label{f3}
\end{figure}
To explore the thermal evolution of these three classes, we refer to Fig. 3, where each class exhibits a specific thermal behavior. For the first class, $W = -1$, represented by the black curve and corresponding to the Tsallis parameter exceeds the critical value, we observe that the event horizon increases as the inverse temperature increases. This indicates that small black holes emit high-temperature Hawking radiation, while large black holes emit low-temperature Hawking radiation. This thermodynamic behavior is consistent with that of Schwarzschild black holes within traditional statistics. Additionally, this class is characterized by a negative topological number $(W = -1)$, which reflects its thermodynamic instability and aligns with the known instability of Schwarzschild black holes in
standard frameworks.
\\

For the second class, $W = +1$, represented by the red curve and corresponding to $\delta < 1/2$, we observe that the event horizon decreases as the inverse temperature of the black hole increases. This
means that black holes with low mass, i.e., small black holes, vaporize by emitting low-temperature Hawking radiation, while large black holes emit high-temperature Hawking radiation. The positive
topological number $(W = +1)$ of this class indicates thermodynamic stability. Notably, this stability introduces a significant modification to the behavior of Schwarzschild black holes, as the adoption of
Tsallis statistics resolves their instability observed in the classical thermodynamic framework. This result emphasizes the influence of non-extensive statistics in altering the thermodynamic properties of
black holes.
\\

For the third class, \(W = 0\), where the Tsallis parameter equals its critical value, represented by the green curve, we observe that the inverse temperature remains constant across all values of the event horizon. This implies that the temperature of the Hawking radiation emitted by black holes is independent of their size, mass, or entropy. On the other hand, the behavior of charged black holes in Anti-de Sitter  spacetime closely resembles that of the Van der Waals fluid \cite{11}, where the cosmological constant is interpreted as a thermodynamic variable and associated with thermodynamic pressure \cite{10}. As a result, three distinct phases emerge for charged  black holes: small, medium, and large black holes mirroring the classical phases of a Van der Waals fluid. The temperature of small black holes increases with the event horizon, medium black holes exhibit thermodynamic instability, and large black holes display nearly linear thermal evolution with a very small slope, often considered quasi-static. Consequently, the thermal behavior of Schwarzschild black holes in the class \(W = 0\) is similar to that of large charged black holes in AdS spacetime.  Furthermore, this class represents the transition point between  stable (\(W = +1\)) and unstable (\(W = -1\)) regimes, marking the critical boundary in thermodynamic topology.
\\

These stability behaviors of Schwarzschild black holes can also be supported by analyzing the evolution of their heat capacity within the framework of Tsallis statistics. The heat capacity is defined as
\begin{equation}
	C = T_T \frac{\partial S_T}{\partial T_T} = \frac{2\delta \left(\pi r^2\right)^\delta}{1 - 2\delta}.
\end{equation}

\begin{figure}[htp]
	\centering
	\includegraphics[width=0.5\linewidth]{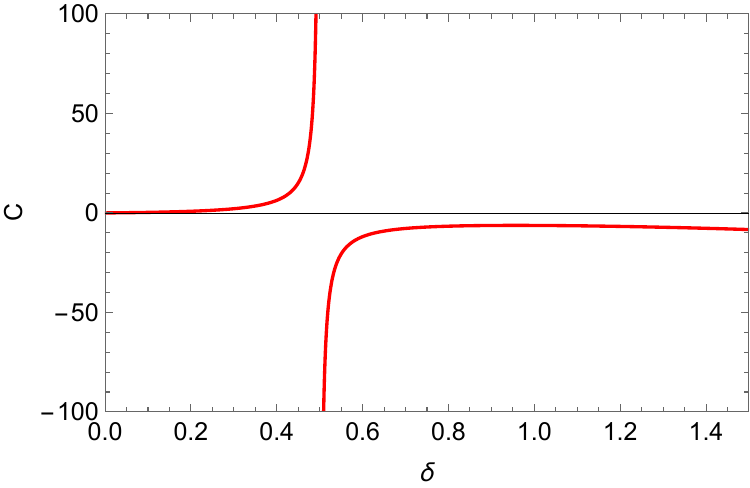}
	\caption{ Heat capacity in terms of the Tsallis parameter with a fixed event horizon of black holes.}
	\label{f4}
\end{figure}
Fig. \ref{f4} illustrates the effect of the Tsallis parameter on the heat capacity of Schwarzschild black holes.  For $\delta < 1/2$, the heat capacity is positive, indicating a stable phase. Conversely, for $\delta > 1/2$, the heat capacity becomes negative, signifying an unstable phase. At $\delta = 1/2$, the heat capacity reaches a critical point, marking a phase transition between stable and unstable phases. These findings further support the conclusions drawn from thermodynamic topology.
\subsection{Schwarzschild and Large Charged AdS Black Holes}
We analyze the similarity between Schwarzschild black holes in class $W=0$ and large charged AdS black holes using topological thermodynamics within Tsallis statistics. The metric function for charged AdS black holes is given by
\begin{equation}
 f(r) = 1 - \frac{2M}{r} + \frac{Q^2}{r^2} + \frac{r^2}{l^2},
\end{equation}
where $l$ is the AdS radius, related to the thermodynamic pressure $P$ by
\begin{equation}
 P = \frac{3}{8\pi l^2}.
\end{equation}
The mass of the black hole is expressed as
\begin{equation}
 M = \frac{3 r_+^2 + 3 Q^2 + 8\pi P r_+^4}{6 r_+},
\end{equation}
where $r_+$ is the event horizon of black hole. The off-shell free energy reads
\begin{equation}
 \mathcal{F}_T = M - \frac{S_T}{\tau}
 = \frac{3 r_+^2 + 3 Q^2 + 8\pi P r_+^4}{6 r_+} - \frac{(\pi r_+^2)^\delta}{\tau}.
\end{equation}
The vector field components are
\begin{align}
 \phi^1 &= \frac{\partial\mathcal{F}_T}{\partial r_+} = \frac{r_+}{2} + \frac{Q^2}{2r_+} - \frac{(\pi r_+^2)^\delta}{\tau} + 4\pi P r_+^4, \\
 \phi^2 &= -\cot\theta\,\csc\theta.
\end{align}
We parametrize the inverse temperature as
\begin{equation}
 \tau = \frac{4 r_0 (\pi r_0^2)^\delta \delta}{r_0^2 - Q^2 + 8\pi P r_0^4},
\end{equation}
where $r_0$ is an arbitrary  radius with the dimension of length. When $r_0 = r_+$, this inverse temperature equals the inverse temperature of the charged AdS black hole in Tsallis statistics, corresponding to the zero point of the vector field.

Here, we compare the thermodynamic behavior of Schwarzschild and large charged AdS black holes in Tsallis statistics. large AdS black holes are characterized by $r_+ > r_0$. We fix $Q = P = 1$. 

\begin{figure}[htp]
    \centering
     \begin{subfigure}{0.3\textwidth}
        \centering
        \includegraphics[width=\linewidth]{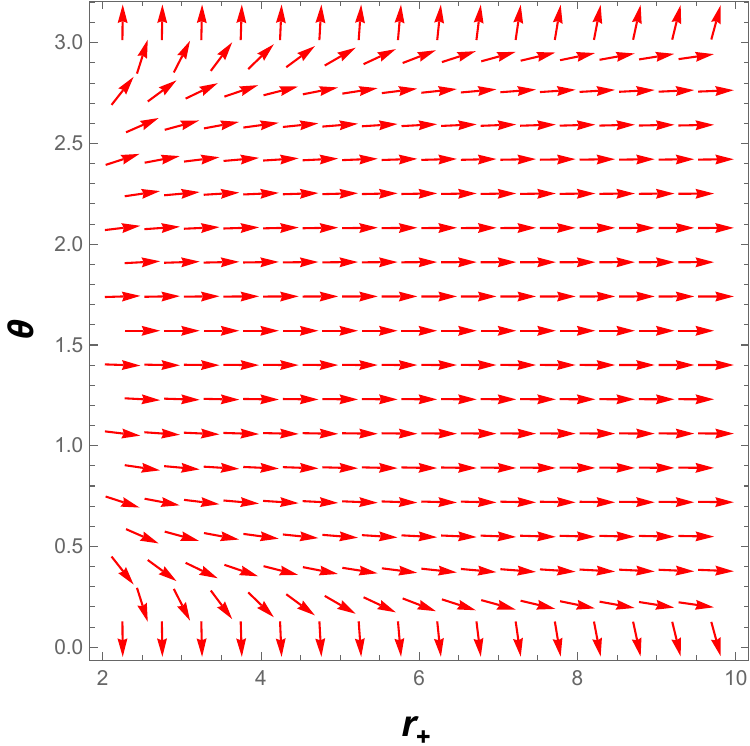}
        \caption{$\delta = 0.4$}
        \label{ad3}
    \end{subfigure}
    \begin{subfigure}{0.3\textwidth}
        \centering
        \includegraphics[width=\linewidth]{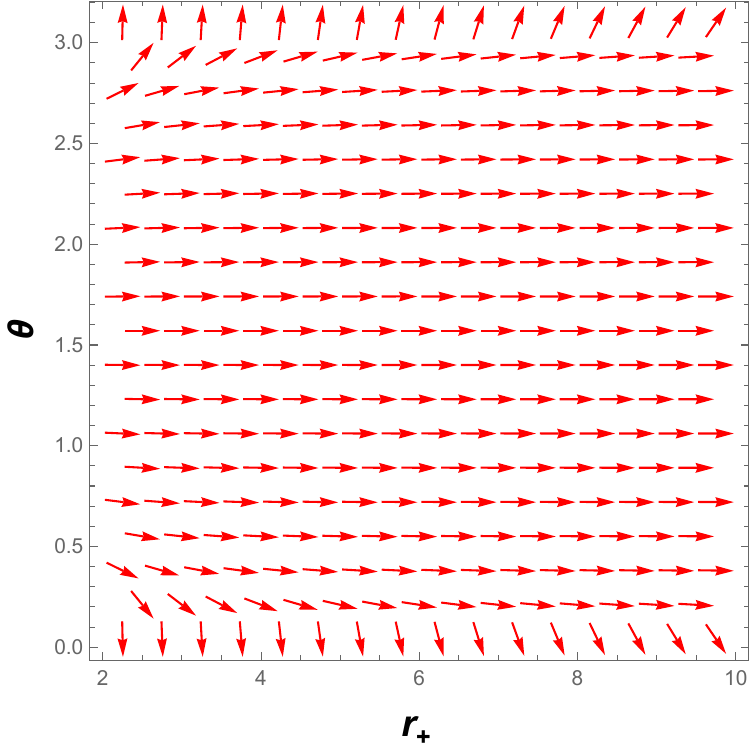}
        \caption{$\delta = 0.8$}
        \label{ad1}
    \end{subfigure}
    \begin{subfigure}{0.3\textwidth}
        \centering
        \includegraphics[width=\linewidth]{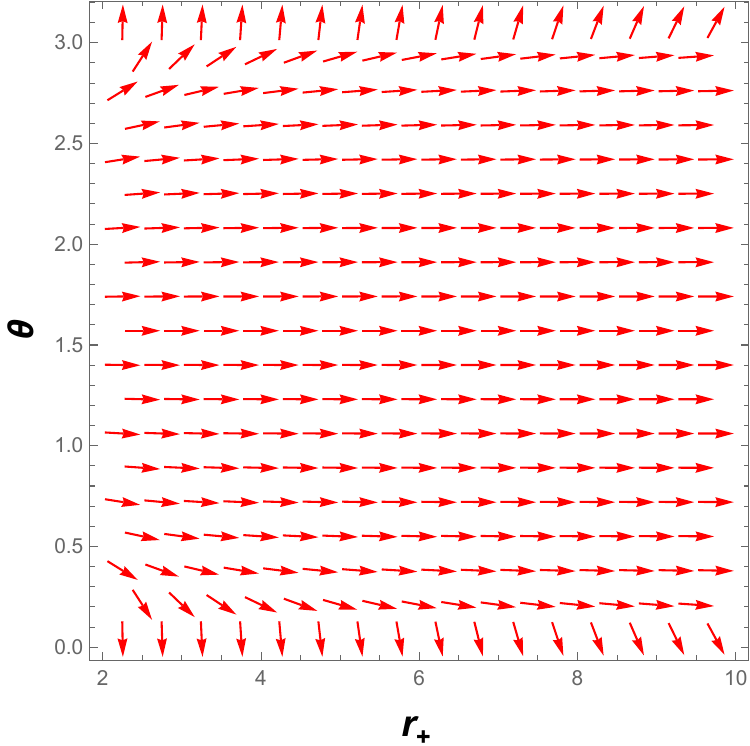}
        \caption{$\delta = 1.2$}
        \label{ad2}
    \end{subfigure}
    \caption{Vector field $\phi$ of large charged AdS black holes for different Tsallis parameters in the $\theta$–$r_+$ plane, with $r_0 = 1$.}
    \label{AdSc}
\end{figure}

Fig.~\ref{AdSc} shows the vector field $\phi$ for large charged AdS black holes ($r_+ > r_0$). We observe that the topological thermodynamics of these black holes is unaffected by the Tsallis parameter, and no zero point appears, indicating that their topological number is $W=0$. This behavior is similar to that of Schwarzschild black holes in class $W=0$ (see Fig.~\ref{fig:b}). Therefore, Schwarzschild black holes in class $W=0$ and large charged AdS black holes exhibit the same thermodynamic behavior within the framework of Tsallis statistics.
\subsection{Reissner–Nordström Black Holes}
We investigate the topological classification and thermodynamic stability of Reissner–Nordström  (RN) black holes within Tsallis statistics. The metric function for RN black holes can be written as
\begin{equation}
	f(r) = 1 - \frac{2M}{r} + \frac{Q^2}{r^2}.
\end{equation}
The mass of  RN black hole in terms of the event horizon \(r_+\) is obtained by setting \(f(r_+) = 0\), which yields
\begin{equation}
	M = \frac{r_{+}}{2} + \frac{Q^2}{2r_{+}}.
\end{equation}
The off-shell free energy of the RN black hole is given by
\begin{equation}
	\mathcal{F}_T = M - \frac{S_T}{\tau} = \frac{r_{+}}{2} + \frac{Q^2}{2r_{+}} - \frac{\left(\pi r_+^2\right)^\delta}{\tau}.
\end{equation}
The components of the vector field, derived from the off-shell free energy, are
\begin{equation}
 \label{yhh}
	\phi^1 = \frac{\partial \mathcal{F}_T}{\partial r_+} = \frac{1}{2} - \frac{Q^2}{2r_{+}^2} - \frac{2\,\delta\,(\pi r_+^2)^\delta}{r_+\,\tau},
\end{equation}
and
\begin{equation}
	\phi^2 = -\cot\theta\,\csc\theta.
\end{equation}

To simplify our analysis, we parameterize the inverse temperature as follows
\begin{equation}
	\tau = \frac{4 \delta \, r_0 \left(\pi r_0^2\right)^{\delta}}{r_0^2 - Q^2},
\end{equation}
where \(r_0\) is an arbitrary parameter with the dimension of the  length. For \(r_0 = r_+\), the inverse temperature corresponds to that of the RN black hole, and this point is a zero of \(\phi^1\) (i.e., \(\phi^1 = 0\)). This implies that for all values of the Tsallis parameter, there exists a solution for \(\phi = 0\) at \(r_0 = r_+\) and \(\theta = \pi/2\).

To study the classification and stability of RN black holes, we calculate the winding numbers of all zero points and thereby determine the topological number. Since the winding numbers and zero points cannot be obtained analytically, we employ a numerical method to solve \(\phi = 0\) for different values of the Tsallis parameter, setting \(Q = 1\) and \(r_0 = 3\).

In the first case, \(\delta = 0.4\), we find a single zero point, denoted \(ZP1\), located at \(r_+ = r_0\) and \(\theta = \pi/2\) (see Fig. \ref{rn4}). According to Eq. \eqref{wn}, the winding number is \(w = +1\) and the topological number is \(W = +1\), which indicates that the RN black holes are both locally and globally stable in this case.

For the second case, \(\delta = 0.8\), we find two zero points, denoted \(ZP2\) and \(ZP3\), as shown in Fig. \ref{rn8}. Using Eq. \eqref{wn}, we obtain a winding number \(w_2 = +1\) for \(ZP2\) and \(w_3 = -1\) for \(ZP3\). Consequently, the topological charge is given by
\begin{equation}
	W = w_2 + w_3 = 0.
\end{equation}
We conclude that RN black holes are stable at \(ZP2\) and unstable at \(ZP3\); in other words, for \(\delta = 0.8\) the small black holes are stable while the large black holes are unstable. Moreover, in terms of global stability, the RN black holes in this case are less stable than those in the \(\delta = 0.4\) case.

For the cases presented in Figs. \ref{rn12} and \ref{rn16}, we obtain results analogous to those for \(\delta = 0.8\): small black holes  stable and large black holes are unstable. Overall,  RN black holes for these values of the Tsallis parameter are less stable than those at \(\delta = 0.4\).

\begin{figure}[htp]
	\centering
	\includegraphics[width=0.4\linewidth]{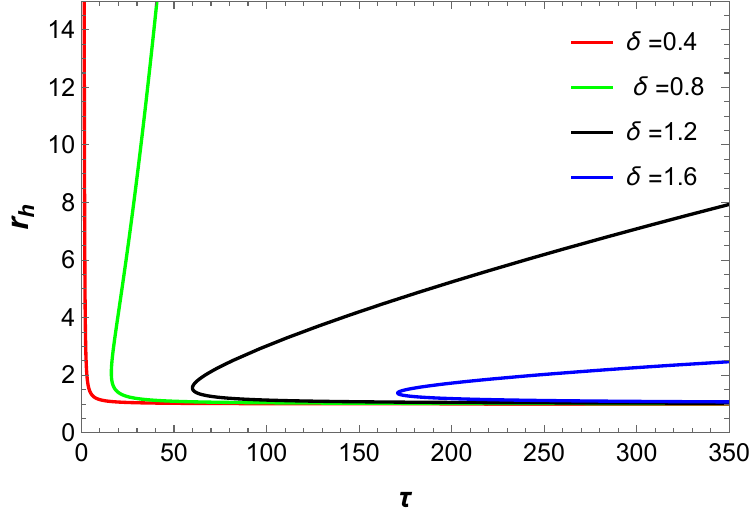}
	\caption{Zero-point curves of the vector component \(\phi^1\) in the \(\tau\)-\(r_+\) plane for different values of the Tsallis parameter.}
	\label{fa33}
\end{figure}

\begin{figure}[ht]
	\centering
	\begin{subfigure}{0.33\textwidth}
		\centering
		\includegraphics[width=\linewidth]{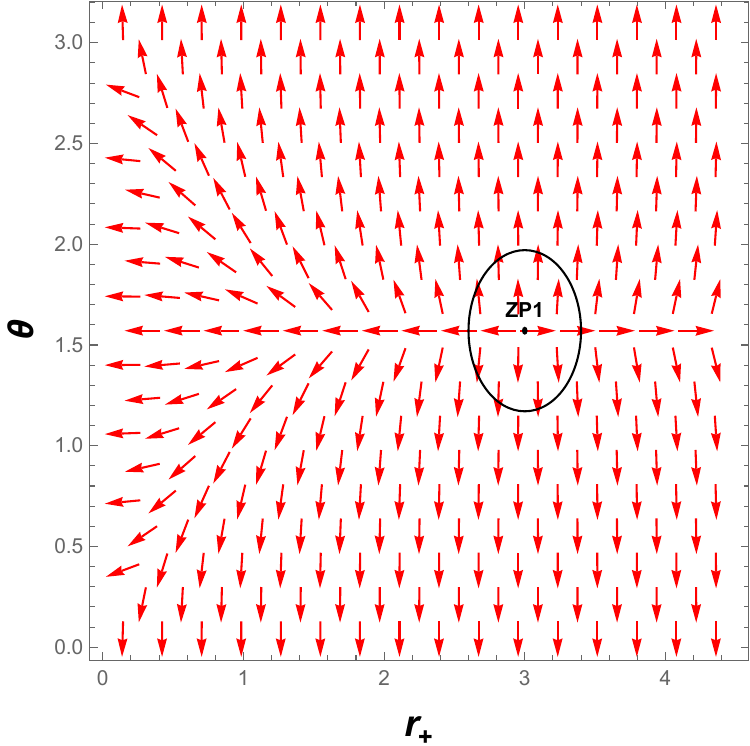}
		\caption{\(\delta = 0.4\)}
		\label{rn4}
	\end{subfigure}
	\begin{subfigure}{0.33\textwidth}
		\centering
		\includegraphics[width=\linewidth]{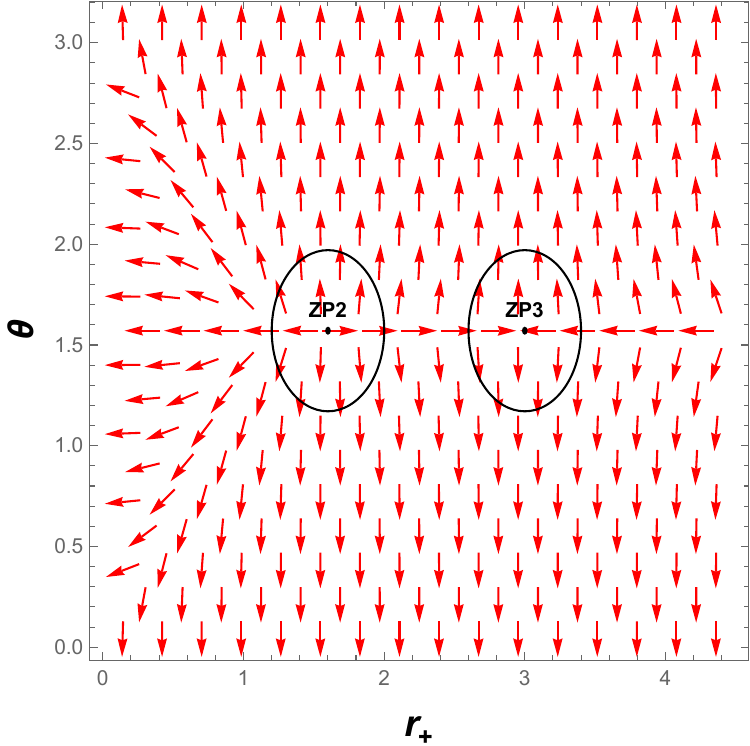}
		\caption{\(\delta = 0.8\)}
		\label{rn8}
	\end{subfigure}
	\begin{subfigure}{0.33\textwidth}
		\centering
		\includegraphics[width=\linewidth]{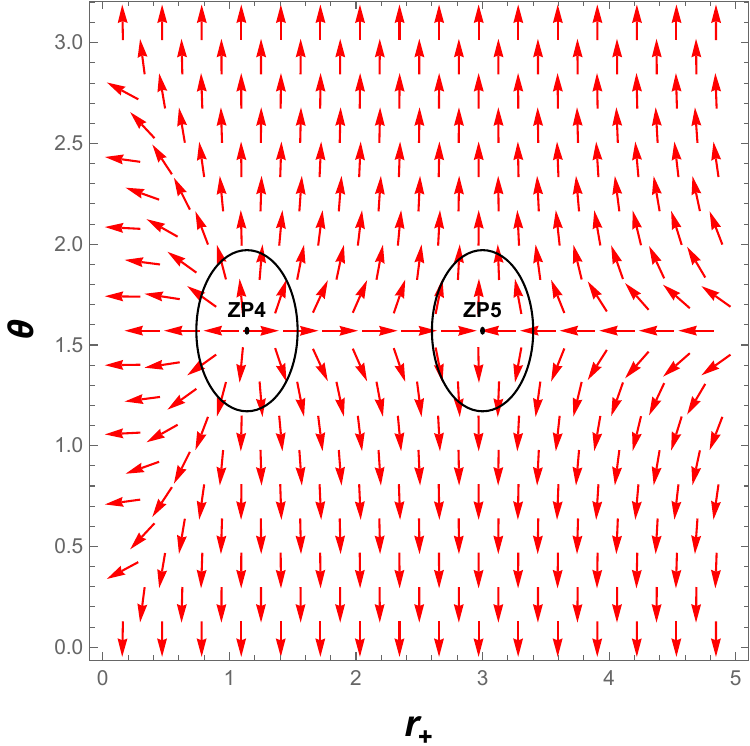}
		\caption{\(\delta = 1.2\)}
		\label{rn12}
	\end{subfigure}
	\begin{subfigure}{0.33\textwidth}
		\centering
		\includegraphics[width=\linewidth]{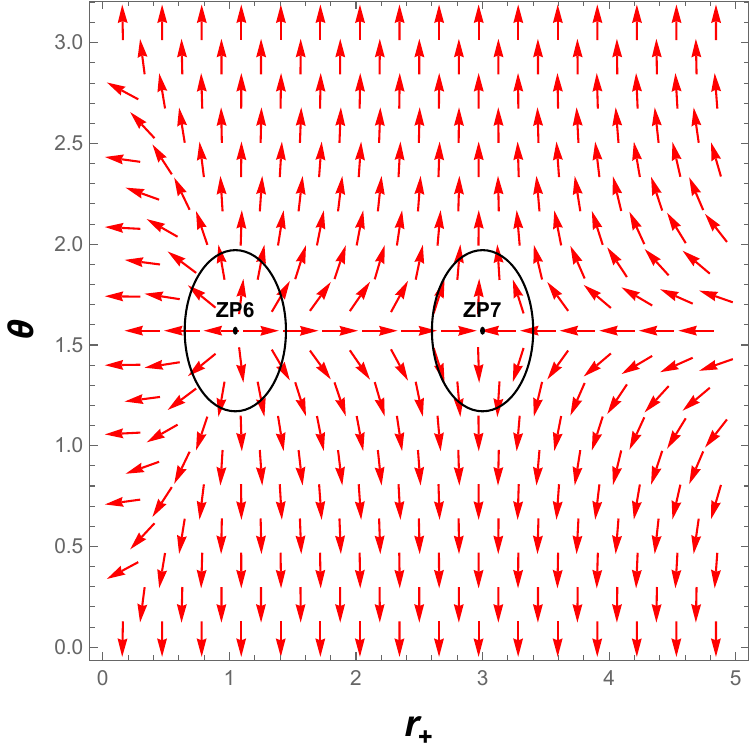}
		\caption{\(\delta = 1.6\)}
		\label{rn16}
	\end{subfigure}
	\caption{Vector field \(\phi\) for RN black holes for different values of the Tsallis parameter in the \(\theta\)-\(r_+\) plane, with \(r_0 = 3\) and \(Q = 1\).}
	\label{frn}
\end{figure}

In conclusion, we classify  RN black holes within Tsallis statistics into two classes characterized by topological numbers \(W = +1\) and \(W = 0\). To explore the thermal behavior of these classes, Fig. \ref{fa33} shows the evolution of the event horizon as a function of the inverse temperature for different values of the Tsallis parameter. For the thermal evolution of the first class (\(W = +1\)), we observe behavior similar to that of Schwarzschild black holes (see the red curve in Fig. \ref{f3}), where the event horizon decreases as the inverse temperature increases. This behavior, corresponding to a positive topological number, indicates a stable thermodynamic phase. Thus, at small values of the Tsallis parameter, the presence or absence of electric charge does not affect the stability of the black holes.

For the second class (\(W = 0\)), Fig. \ref{fa33} reveals two branches. The upper branch corresponds to large black holes, which are characterized by a negative winding number and are therefore unstable, while the lower branch corresponds to small black holes, characterized by a positive winding number and hence stable. Notably, this phenomenon is related to the presence of electric charge in the black holes—a feature absent in Schwarzschild black holes.
  
\subsection{Charged and Non-Charged 5d Black Holes}
 We explore the classification and thermodynamic behavior of both charged and non-charged black holes in five dimensions within the context of Tsallis statistics. We begin with charged 5d black holes, which are described by the metric function
\begin{equation}
	f(r) = 1 - \frac{8 M}{3 \pi r^2} + \frac{4 Q^2}{3 \pi^2 r^4}.
\end{equation}
For \(Q=0\), we recover the solution for non-charged black holes. By solving \(f(r_+) = 0\), the black hole mass is given by
\begin{equation}
	M = \frac{Q^2}{2\pi r_+^2} + \frac{3 \pi r_+^2}{8}.
\end{equation}
The Tsallis entropy of charged black holes in 5 dimensions is defined as
\begin{equation}
	S_T = \left(\frac{\pi^2 r_+^3}{2}\right)^\delta.
\end{equation}
Additionally, the off-shell free energy is expressed as
\begin{equation}
	\mathcal{F}_T = M - \frac{S_T}{\tau} = \frac{3 \pi r_+^2}{8} + \frac{Q^2}{2\pi r_+^2} - \frac{\left(\frac{\pi^2 r_+^3}{2}\right)^\delta}{\tau}.
\end{equation}

The components of the field vector are given by
\begin{equation}
	\phi^1 = \frac{\partial \mathcal{F}_T}{\partial r_+} = \frac{3 \pi r_+}{4} - \frac{Q^2}{\pi r_+^3} - \frac{3\delta \left(\frac{\pi^2 r_+^3}{2}\right)^\delta}{\tau\,r_+},
\end{equation}
and
\begin{equation}
	\phi^2 = -\cot\theta\,\csc\theta.
\end{equation}

To simplify our study, we parameterize the inverse temperature \(\tau\) using an arbitrary parameter \(r_0\) (with the dimension of length) as follows
\begin{equation}
	\tau = \frac{12\pi\,\delta\, r_0^2 \left(\frac{\pi^2 r_0^3}{2}\right)^\delta}{3\pi^2\, r_0^4 - 4 Q^2}.
\end{equation}
where \(r_0\) is an arbitrary parameter with the dimension of length. For \(r_0 = r_+\), the inverse temperature corresponds to that of the black hole, and \(\phi^1 = 0\) at this point. By solving \(\phi = 0\), we find a single zero point at \(r_+ = r_0\) and \(\theta = \pi/2\).

\begin{figure}[htp]
	\centering
	\begin{subfigure}{0.33\textwidth}
		\centering
		\includegraphics[width=\linewidth]{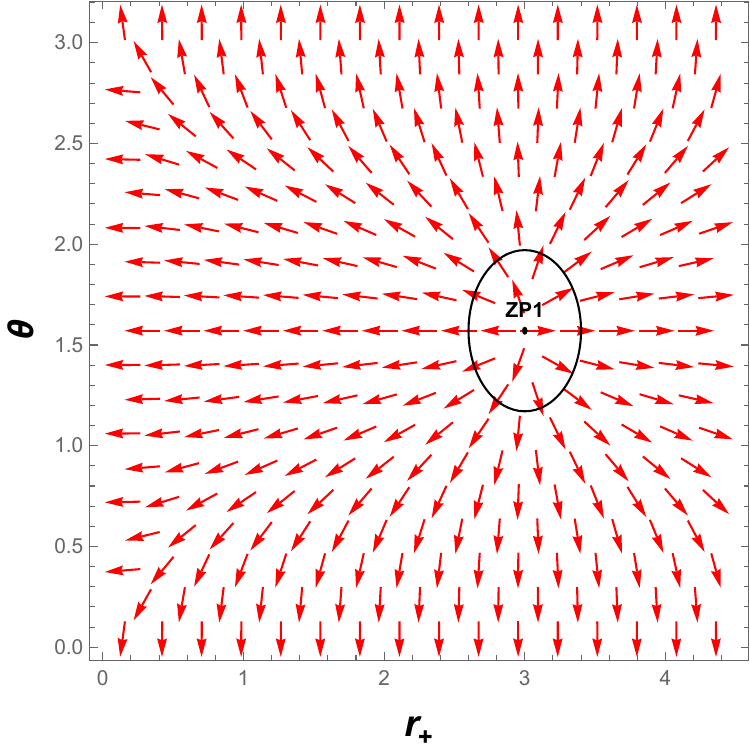}
		\caption{\(\delta = 0.4\)}
		\label{hc4}
	\end{subfigure}
	\begin{subfigure}{0.33\textwidth}
		\centering
		\includegraphics[width=\linewidth]{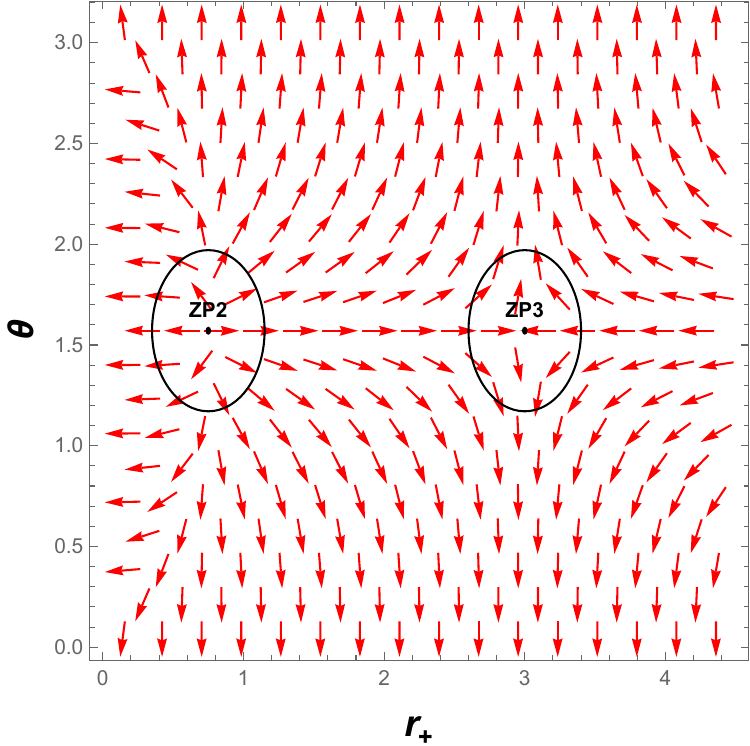}
		\caption{\(\delta = 0.8\)}
		\label{hc8}
	\end{subfigure}
	\begin{subfigure}{0.33\textwidth}
		\centering
		\includegraphics[width=\linewidth]{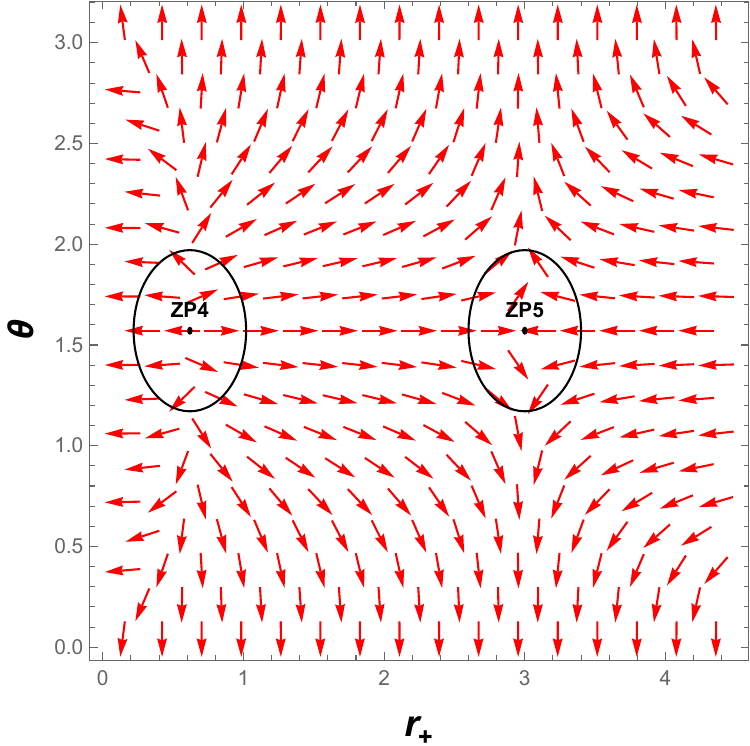}
		\caption{\(\delta = 1.2\)}
		\label{hc12}
	\end{subfigure}
	\begin{subfigure}{0.33\textwidth}
		\centering
		\includegraphics[width=\linewidth]{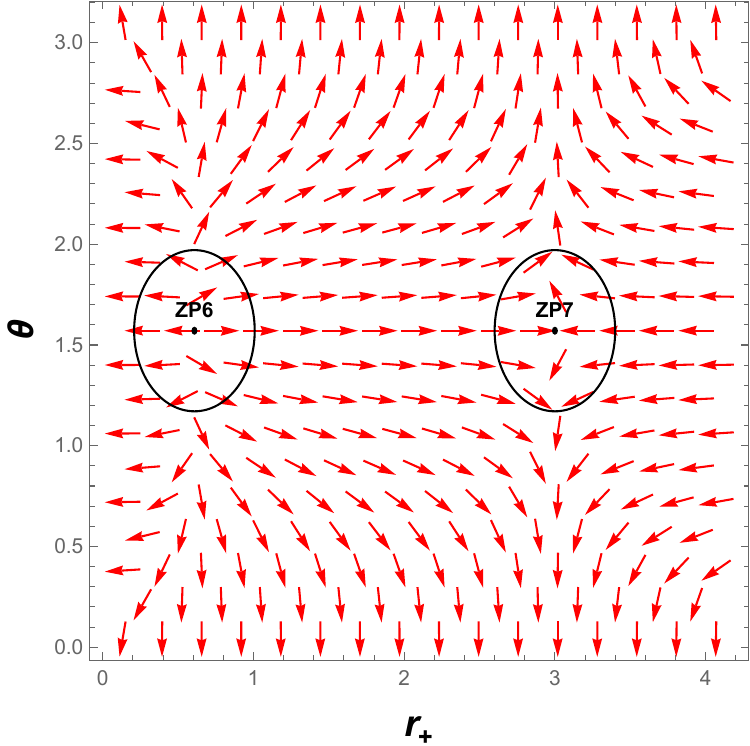}
		\caption{\(\delta = 1.6\)}
		\label{hc16}
	\end{subfigure}
	\caption{Vector field \(\phi\) of charged 5d black holes for different values of the Tsallis parameter in the \(\theta\)-\(r_+\) plane, with \(r_0 = 3\) and \(Q = 1\).}
	\label{frn5d}
\end{figure}

Since the zero points and winding numbers cannot be obtained analytically for these black holes, we employ numerical methods by fixing \(Q = 1\) and \(r_0 = 3\). The zero points are illustrated in Fig. \ref{frn5d} for different values of the Tsallis parameter. For \(\delta = 0.4\), there is a single zero point, denoted \(ZP1\), as shown in Fig. \ref{hc4}. Using Eq. \eqref{wn}, the winding number at \(ZP1\) is found to be \(w_1 = +1\), indicating local stability; its topological number is \(W = +1\), corresponding to global stability.

\begin{figure}[htp]
	\centering
	\begin{subfigure}{0.405\textwidth}
		\centering
		\includegraphics[width=\linewidth]{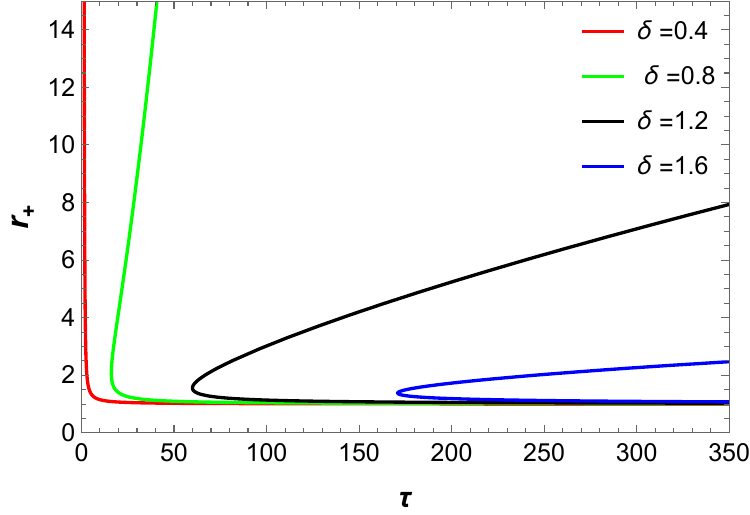}
		\caption{Charged black holes }
		\label{ftch5d}
	\end{subfigure}
	\begin{subfigure}{0.4\textwidth}
		\centering
		\includegraphics[width=\linewidth]{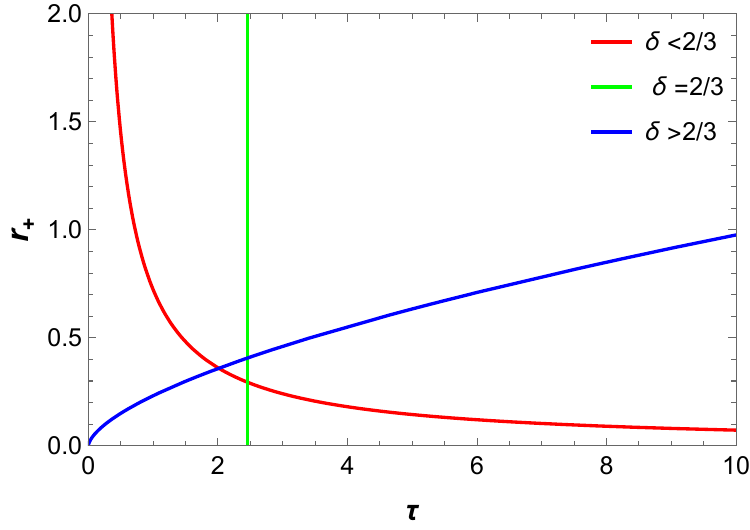}
		\caption{Non-charged black holes}
		\label{ftrn}
	\end{subfigure}
	\caption{Zero point curves of the vector \(\phi^1\) in the \(\tau\)-\(r_h\) plane for different values of the Tsallis parameter.}
	\label{yahya}
\end{figure}

For the other cases (\(\delta = 0.8,\, 1.2,\, 1.6\)), as shown in Figs. \ref{hc8}, \ref{hc12}, and \ref{hc16}, we observe two zero points. In these cases, the smaller zero point has a positive winding number (indicating local stability for small black holes), while the larger zero point has a negative winding number (indicating local instability for large black holes). Consequently, the topological number for these cases is \(W  = 0\), implying that this class is globally less stable than the \(W = +1\) class.
The thermal evolution of these classes is depicted in Fig. \ref{ftch5d}. For the first class (\(W = +1\)), the event horizon decreases as the inverse temperature increases, corresponding to a stable thermodynamic phase. For the second class (\(W = 0\)), two branches are observed: the upper branch corresponds to large (unstable) black holes, while the lower branch corresponds to small (stable) black holes. These classifications and thermodynamic behaviors are similar to those of charged black holes in 4 dimensions, indicating that the number of dimensions does not significantly affect the topological thermodynamics.
\\

\begin{figure}[htp]
	\centering
	\begin{subfigure}{0.3\textwidth}
		\centering
		\includegraphics[width=\linewidth]{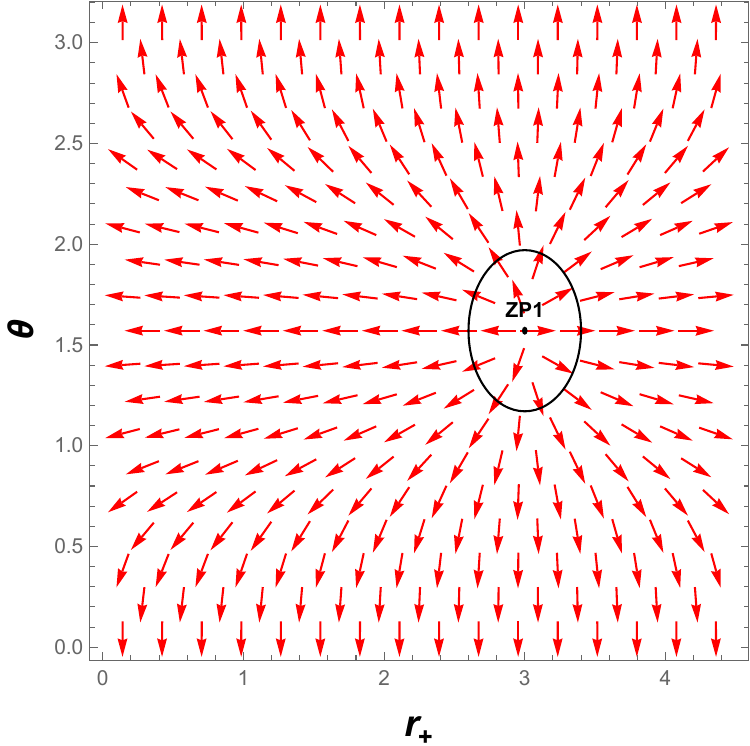}
		\caption{\(\delta < 2/3\)}
		\label{hrn4}
	\end{subfigure}
	\begin{subfigure}{0.3\textwidth}
		\centering
		\includegraphics[width=\linewidth]{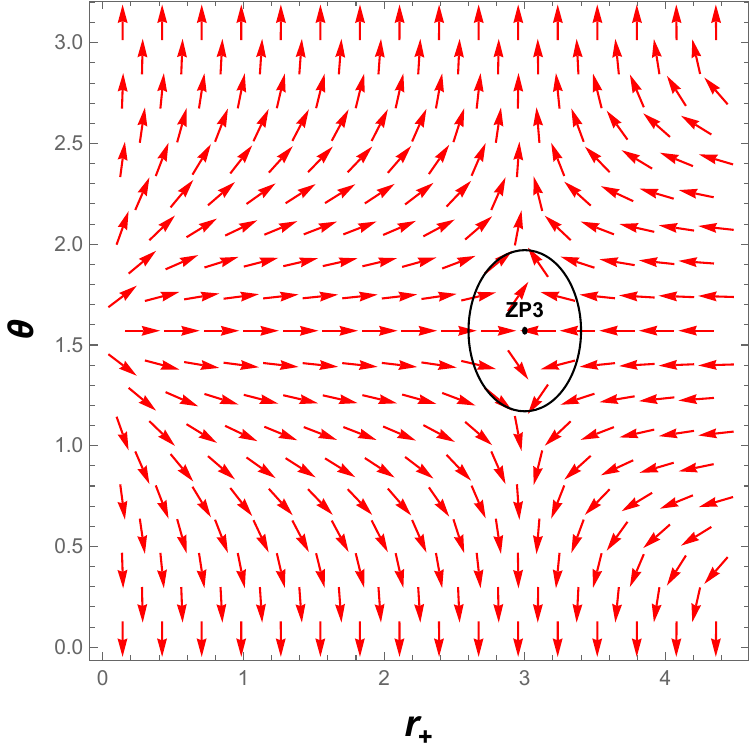}
		\caption{\(\delta > 2/3\)}
		\label{hrn12}
	\end{subfigure}
	\caption{Vector field \(\phi\) of non-charged 5d black holes for different values of the Tsallis parameter in the \(\theta\)-\(r_+\) plane, with \(r_0 = 3\).}
	\label{frnc}
\end{figure}

%***********************************************
For the non-charged black holes (\(Q=0\)), by solving \(\phi = 0\) we find a single zero point at \(r_+ = r_0\) and \(\theta = \pi/2\). Furthermore, using Eq. \eqref{yhh}, the winding number is determined by 
\begin{equation}
	w = \operatorname{sign}\!\left[\left.\frac{\partial \phi^1}{\partial r_+}\right|_{r_+=r_0}\right] = \operatorname{sign}\!\left[\frac{3\pi (2 - 3\delta)}{4}\right].
	\label{tt}
\end{equation}
Eq. \eqref{tt} and Fig. \ref{frnc} show that the winding number depends on the value of the Tsallis parameter. Specifically, when \(\delta < 2/3\) the winding number is \(w = +1\), indicating local stability, whereas for \(\delta > 2/3\) the winding number becomes \(w = -1\), indicating local instability. The case \(w = 0\) at \(\delta = 2/3\) represents a phase transition point and the critical value of the Tsallis parameter.

Using the topological number, we classify these black holes into three classes: the first class with \(W = +1\) for \(\delta < 2/3\) (stable), the second class with \(W = -1\) for \(\delta > 2/3\) (unstable), and the third class with \(W = 0\) at \(\delta = 2/3\) (critical).

Fig. \ref{ftrn} illustrates the thermal evolution of the black holes for each classes. For the first class (\(W = +1\), red curve), the event horizon decreases as the inverse temperature increases. For the second class (\(W = -1\), blue curve), the event horizon increases with increasing inverse temperature. For the third class (\(W = 0\), green curve), corresponding to the critical value \(\delta = 2/3\), the inverse temperature remains constant as the event horizon varies. These thermodynamic behaviors are analogous to those of 4-dimensional Schwarzschild black holes, with the primary difference being the value of the critical Tsallis parameter. Consequently, we conclude that the number of dimensions does not significantly impact the thermodynamic topology of non-charged black holes.
\section{Higher-Dimensional Black Holes}
\label{SS3}
We extend the topological thermodynamics analysis to charged and uncharged black holes in $d$ spacetime dimensions using Tsallis statistics. We classify their topological properties, examine thermodynamic stability, and assess the role of dimensionality.
The general static, spherically symmetric metric in $d$  dimensions reads
\begin{equation}
 ds^2 = -f(r)\,dt^2 + \frac{dr^2}{f(r)} + r^2\,d\Omega_{d-2}^2,
\end{equation}
where $d\Omega_{d-2}^2$ denotes the line element on the unit $(d-2)$-sphere. The metric function is
\begin{equation}
\label{gmt}
 f(r) = 1 - \frac{2m}{r^{d-3}} + \frac{q^2}{r^{2(d-3)}},
\end{equation}
with parameters $m$ and $q$ related to the  mass, $M$, and charge, $Q$, by
\begin{equation}
 M = \frac{d-2}{8\pi}\,\omega_{d-2}\,m,
 \quad Q = \frac{\sqrt{2(d-2)(d-3)}}{8\pi}\,\omega_{d-2}\,q,
\end{equation}
where
\begin{equation}
 \omega_{d-2} = \frac{2\pi^{(d-1)/2}}{\Gamma\bigl((d-1)/2\bigr)}.
\end{equation}
The Bekenstein–Hawking entropy is
\begin{equation}
 S_{\rm BH} = \frac{\omega_{d-2}\,r_+^{d-2}}{4},
\end{equation}
where $r_+$ is the event-horizon radius. In Tsallis statistics, the entropy writes
\begin{equation}
 S_T = \bigl(S_{\rm BH}\bigr)^\delta = \biggl(\frac{\omega_{d-2}\,r_+^{d-2}}{4}\biggr)^\delta,
\end{equation}
\subsection{Non‐charged Higher‐Dimensional Black Holes}

In this subsection, we focus on non‐charged higher‐dimensional black holes, where we study the topological classification and the impact of the spacetime dimension on the thermodynamic properties within the framework of Tsallis statistics.
\\
To analyze the local and global stability of these black holes, we need expressions for various thermodynamic quantities such as the mass, temperature, and free energy. Using the general metric, Eq.~\eqref{gmt}, the black hole mass is 
\begin{equation}
    M = \frac{d-2}{16\pi}\,r_+^{\,d-3}\,\omega_{d-2}\,.
\end{equation}
The Tsallis temperature of these black holes is then given by
\begin{equation}
    T_T = \frac{\partial M}{\partial S_T}
        = \frac{4^{-2+\delta}\,(d-3)\,\bigl(r_+^{\,d-2}\,\omega_{d-2}\bigr)^{\,1-\delta}}
               {\pi\,r_+\,\delta}\,.
\end{equation}
The off‐shell free energy reads
\begin{equation}
   \mathcal{F}_T \;=\; M \;-\;\frac{S_T}{\tau}
    = \frac{(d-2)\,r_+^{\,d-3}\,\omega_{d-2}}{16\pi}
      \;-\;\frac{4^{-\delta}\,\bigl(r_+^{\,d-2}\,\omega_{d-2}\bigr)^{\!\delta}}{\tau}\,.
\end{equation}
Accordingly, the components of the vector field derived from the off‐shell free energy are
\begin{equation}
    \phi^1 \;=\; \frac{\partial \mathcal{F}_T}{\partial r_+}
      = \frac{d-2}{16\,r_+^4}\,\biggl[
          \frac{(d-3)\,r_+^d\,\omega_{d-2}}{\pi}
          \;-\;\frac{4^{2-\delta}\,\delta\,r_+^3\,\bigl(r_+^{\,d-2}\,\omega_{d-2}\bigr)^{\!\delta}}{\tau}
        \biggr],
\end{equation}
\begin{equation}
    \phi^2 \;=\; -\cot\theta\,\csc\theta\,.
\end{equation}
For convenience, we parametrize the inverse temperature $\tau$ as
\begin{equation}
    \tau = \frac{4^{2-\delta}\,\pi\,r_0\,
                  \bigl(r_0^{\,d-2}\,\omega_{d-2}\bigr)^{-1+\delta}\,\delta}
                 {d-3}\,,
\end{equation}
where $r_0$ is a free parameter with dimensions of length. Note that when $r_0 = r_+$, the inverse temperature, $\tau$, coincides with the inverse temperature of black hole and $\phi^1=0$ at $r_0 = r_+$. Consequently, the vector field has a single zero point of the vector field at
\begin{equation}
    r_+ = r_0 , \qquad \qquad \theta = \frac{\pi}{2}.
\end{equation}
These coordinates are identical for both four-dimensional and higher-dimensional uncharged black holes, showing that the spacetime dimension does not affect the location of the zero point of the vector field. To study how the dimension influences global and local stability, we next compute the winding and topological numbers as follows    
\begin{equation}
\label{wd}
    w = \operatorname{sign}\!\left[\left.\frac{\partial \phi^1}{\partial r_+}\right|_{r_+=r_0}\right] = \operatorname{sign}\!\left[-3 + d + \delta\left(2 - d\right)\right].
\end{equation}
Through the winding number expression in Eq.~\eqref{wd}, we see that \(w\) depends on both the spacetime dimension \(d\) and the nonextensive parameter \(\delta\). Defining the critical Tsallis parameter as 

\begin{equation}
  \delta_c = \frac{d-3}{d-2}\,,
\end{equation}
we find that \(w=+1\) when \(\delta<\delta_c\), \(w=0\) when \(\delta=\delta_c\), and \(w=-1\) when \(\delta>\delta_c\). A positive winding number signals local thermodynamic stability, while a negative winding number indicates local instability; \(w=0\) marks a phase‐transition point. Since there is only one zero point of the vector field for these uncharged black holes, the topological number \(W\) coincides with \(w\). Therefore the black holes are globally stable for \(\delta<\delta_c\) (\(W=+1\)), globally unstable for \(\delta>\delta_c\) (\(W=-1\)), and lie at the stability–instability transition when \(\delta=\delta_c\) (\(W=0\)).
\\

We observe that higher-dimensional  uncharged black holes across all dimensions have the same topological classifications, the only dimension-dependent quantity being the critical Tsallis parameter \(\delta_c\). Consequently, the topological classification of higher-dimensional uncharged black holes within non-extensive statistics is universal.
\subsection{Charged Higher‐Dimensional Black Holes}
We investigate the thermodynamic topology of higher-dimensional black holes within Tsallis statistics, and study the impact of the number of dimensions on topological classification. Through the metric, Eq. \eqref{gmt}, we found the mass of these black holes as follows
\begin{equation}
  M \;= \;  \frac{(d-2)\,r_+^{d-3}\,w_{d-2}}{16\pi}   \;+\; \frac{2\pi Q^2 \,r_+^{3-d}}{(d-3)\,w_{d-2}}.
\end{equation}
Also, the temperature of higher-dimensional black holes within Tsallis statistics is given 
\begin{equation}
 T_T \;=\; \frac{\partial M}{\partial S_T}
        = 
\frac{4^{-2+\delta}\,\bigl(r_+^{\,d-2}\,w_{d-2}\bigr)^{-1-\delta}\,\bigl( (d-3)(d-2)\,r_+^{2d}\,w_{d-2}^2 -32\pi^2 Q^2 r_+^6  \bigr)}{(d-2)\,\pi\,r_+^5\,\delta}.
\end{equation}
To explore the topological classification of these black holes, we determine the off-shell free energy as follows
\begin{equation}
     \mathcal{F}_T \;=\; M \;-\;\frac{S_T}{\tau} = 
\; \frac{(d-2)\,r_+^{d-3}\,w_{d-2}}{16\pi}
\;-\;
\frac{4^{-\delta}\,\bigl(r_+^{d-2}\,w_{d-2}\bigr)^{\delta}}{\tau} \;+\; \frac{r_+^{3-d}\,2\pi\,Q^2}{(d-3)\,w_{d-2}}.
\end{equation}
The components of the field vector are given by
\begin{equation}
    \phi^1  \;=\; \frac{\partial \mathcal{F}_T}{\partial r_+}= 
\frac{(d-3)(d-2)\,r_+^{d-4}\,w_{d-2}}{16\pi}
\;-\;
\frac{4^{-\delta}\,(d-2)\,\delta\,\bigl(r_+^{d-2}\,w_{d-2}\bigr)^{\delta}}{r_+\,\tau} -\frac{2\pi Q^2\,r_+^{2-d}}{w_{d-2}},
\end{equation}
and 
\begin{equation}
    \phi^2 \;=\; -\cot\theta\,\csc\theta\,.
\end{equation}
We parameterize the inverse temperature $\tau$ as follows
\begin{equation}
\tau = \frac{4^{-2+\delta}\,\bigl(r_0^{\,d-2}\,w_{d-2}\bigr)^{-1-\delta}\,\bigl(-32\pi^2 Q^2 r_0^6 + (d-3)(d-2)\,r_0^{2d}\,w_{d-2}^2\bigr)}{(d-2)\,\pi\,r_0^5\,\delta}
\end{equation}

where \( r_0 \) is an arbitrary parameter with the dimension of length. At \( r_0 = r_+ \), the inverse temperature becomes equivalent to the black hole inverse temperature, and the free energy \( \mathcal{F}_T \) becomes the on-shell free energy. This case also corresponds to a zero point of the field vector at \( \theta = \pi/2 \).

To study the classification and the stability of these black holes, we use the following expressions for the winding number \( w \) and the topological number \( W \).

\begin{equation}
    w_i = \operatorname{sign}\!\left[\left.\frac{\partial \phi^1}{\partial r_+}\right|_{r_+=r_i}\right], \qquad \qquad  W = \sum_{i=0}^{N} w_i,
\end{equation}

where \( r_i \) is a coordinate of a zero point with \( \theta = \pi/2 \), and \( N \) is the number of zero points. We cannot find the zero points and winding numbers by analytical methods; for this reason, we use numerical approaches, where we fix the number of dimensions and the Tsallis parameter at different values. In this setup, we fix \( r_0 \) and the electrical charge, with \( r_0 = 3 \) and \( Q = 1 \).

\begin{table}[ht]
    \centering
\begin{tabular}{|l|c|c|c|c|c|c|}
\hline$\delta$ & \multicolumn{2}{|c|}{0.4} & \multicolumn{2}{c|}{0.8} & \multicolumn{2}{c|}{1.2} \\
\hline { 6d-RN Black holes } & $r_i$ & $w_i$ & $r_i$ & $w_i$ & $r_i$ & $w_i$ \\
\cline { 2 - 7 } & $3$ & +1 & \begin{tabular}{c}
0.732 \\
3
\end{tabular} & \begin{tabular}{c}
+1 \\
-1
\end{tabular} & \begin{tabular}{c}
0.585 \\
3
\end{tabular} & \begin{tabular}{c}
+1 \\
-1
\end{tabular} \\
\hline$W$ & \multicolumn{2}{|c|}{+1} & \multicolumn{2}{c|}{0} & \multicolumn{2}{c|}{0} \\
\hline
\end{tabular}
\caption{Winding and topological numbers for six-dimensional charged black holes at different $\delta$ values.}
\label{6d}
\end{table}

\begin{table}[ht]
    \centering
\begin{tabular}{|l|c|c|c|c|c|c|}
\hline$\delta$ & \multicolumn{2}{|c|}{0.4} & \multicolumn{2}{c|}{0.8} & \multicolumn{2}{c|}{1.2} \\
\hline { 7d-RN Black holes } & $r_i$ & $w_i$ & $r_i$ & $w_i$ & $r_i$ & $w_i$ \\
\cline { 2 - 7 } & $3$ & +1 & \begin{tabular}{c}
3
\end{tabular} & \begin{tabular}{c}
+1 
\end{tabular} & \begin{tabular}{c}
0.601 \\
3
\end{tabular} & \begin{tabular}{c}
+1 \\
-1
\end{tabular} \\
\hline$W$ & \multicolumn{2}{|c|}{+1} & \multicolumn{2}{c|}{+1} & \multicolumn{2}{c|}{0} \\
\hline
\end{tabular}
\caption{Winding and topological numbers for seven-dimensional charged black holes at different $\delta$ values.}
\label{7d}
\end{table}
\begin{table}[ht]
    \centering
\begin{tabular}{|l|c|c|c|c|c|c|}
\hline$\delta$ & \multicolumn{2}{|c|}{0.4} & \multicolumn{2}{c|}{0.8} & \multicolumn{2}{c|}{1.2} \\
\hline { 8d-RN Black holes } & $r_i$ & $w_i$ & $r_i$ & $w_i$ & $r_i$ & $w_i$ \\
\cline { 2 - 7 } & $3$ & +1 & \begin{tabular}{c}
3
\end{tabular} & \begin{tabular}{c}
+1 
\end{tabular} & \begin{tabular}{c}
0.632 \\
3
\end{tabular} & \begin{tabular}{c}
+1 \\
-1
\end{tabular} \\
\hline$W$ & \multicolumn{2}{|c|}{+1} & \multicolumn{2}{c|}{+1} & \multicolumn{2}{c|}{0} \\
\hline
\end{tabular}
\caption{Winding and topological numbers for eight-dimensional charged black holes at different $\delta$ values.}
\label{8d}
\end{table}

Tables~\eqref{6d}--\eqref{8d} present the values of the event horizons corresponding to the zero points and their associated winding numbers, as well as the topological numbers of charged black holes in six, seven, and eight dimensions for different values of the Tsallis parameter. We find the same classes of black holes in these dimensions, where the classification is determined by the Tsallis parameter. These classes are \( W = +1 \) and \( W = 0 \). For the first class ($W = +1$), a single zero point exists. This class corresponds to small values of the Tsallis parameter and indicates thermodynamic stability.
In the second class ($W = 0$), which is less stable than the first, two zero points appear. The first zero point corresponds to small black holes with a positive winding number, indicating local stability. The second zero point represents large black holes with a negative winding number, associated with local instability.
We conclude that the topology of black holes is not affected by the number of dimensions; it only affects the values of some zero points. Consequently, the topological classification of higher-dimensional charged black holes in the framework of non-extensive statistics is universal.

\section{Conclusions and Discussions}
\label{SS4}
This paper investigates the thermodynamic topology of various black holes, including Schwarzschild, Reissner–Nordström, and both charged and non-charged higher-dimensional black holes, within the framework of Tsallis statistics, a non-extensive generalization of Gibbs–Boltzmann statistics. By considering black holes as topological defects and employing the generalized off-shell free energy along with Duan’s \(\phi\)-mapping theory, we analyze local thermodynamic stability via the winding number \(w\) and global stability through topological numbers (often referred to as topological charges). This approach allows us to classify the black holes into three distinct classes, each characterized by a unique thermodynamic behavior based on the value of the Tsallis parameter. Moreover, we uncover a remarkably unified, dimension agnostic picture of stability and criticality that depends  on the Tsallis parameter \(\delta\). Consequently, the topological classification of higher-dimensional black holes within the framework of non-extensive statistics is universal. 
\\

For Schwarzschild black holes, three classes arise depending on the Tsallis  parameter. When \(\delta > 1/2\), the topological number \(W = -1\) reflects thermodynamic instability, which is consistent with the conventional behavior of Schwarzschild black holes. Conversely, for \(\delta < 1/2\), the topological number \(W = +1\) indicates a transition to a stable thermodynamic phase. At the critical value \(\delta = 1/2\), the topological number \(W = 0\) represents a boundary between stability and instability, with no zero points in the vector field. Moreover, the thermal evolution of these classes reveals the impact of Tsallis statistics on black hole thermodynamics: for the first class (\(W = -1\)), small black holes emit high-temperature Hawking radiation while large black holes emit low-temperature radiation, indicating unstable behavior. In the second class (\(W = +1\)), the behavior is reversed, with low-mass black holes vaporizing through low-temperature radiation and large black holes emitting high-temperature Hawking radiation. In the critical case (\(W = 0\)), the Hawking temperature remains essentially unaffected by  black hole parameters. This latter behavior is reminiscent of large charged black holes in AdS spacetime, whose temperature remains nearly constant. 
Moving beyond four dimensions, the same threefold classification and topological thermodynamic behavior persist for Schwarzschild black holes.
\\

For Reissner–Nordström black holes, the presence of electric charge enriches the topological structure: only two classes emerge (\(W = +1\) and \(W = 0\)). At small values of \(\delta\), the charged black holes behave analogously to the stable branch of Schwarzschild black holes (W=+1). For larger values of \(\delta\), the \(W = 0\) class exhibits two branches: one corresponding to stable (small) black holes and the other corresponding to unstable (large) black holes. 
Furthermore, higher-dimensional charged black holes exhibit the same topological thermodynamic behavior as charged black holes in four-dimensional spacetime.
\\

Our results underscore the pivotal role of non-extensive entropy in modifying black hole stability and phase structure. By tuning the Tsallis parameter, one can induce topological transitions between stable and unstable regimes. Furthermore, the independence of the classification from spacetime dimension suggests a universality of topological thermodynamics in non-extensive frameworks. Several avenues for future research arise from this study. It would be instructive to extend the analysis to Kerr–Newman black holes, as well as to asymptotically AdS spacetimes. Additionally, incorporating other generalized entropies, such as Rényi and Kaniadakis entropies, and exploring their topological impact could further deepen our understanding of black hole thermodynamics.

\section*{Acknowledgments}
Y. Ladghami gratefully acknowledges the support from the "PhD-Associate Scholarship – PASS" grant provided by the National Center for Scientific and Technical Research in Morocco, under grant number 42 UMP2023.

\end{document}